# Synthesis and characterization of HAp nanorods from a cationic surfactant template method

J.M. Coelho, J. Agostinho Moreira, and A. Almeida

*IFIMUP and IN- Institute of Nanoscience and Nanotechnology. Departamento de Física e Astronomia da Faculdade de Ciências da Universidade do Porto. Rua do Campo Alegre, 687. 4169-007 Porto. Portugal.*

F. J. Monteiro

*INEB – Instituto de Engenharia Biomédica, Laboratório de Biomateriais, Rua do Campo Alegre 823, 4150-180 Porto, Portugal.*

*Departamento de Engenharia Metalúrgica e Materiais, Faculdade de Engenharia, Universidade do Porto, Porto, Portugal.*

*Abstract*

**Hydroxyapatite (HAp) [$Ca_{10}(PO_4)_6(OH)_2$] nanorods were synthesized using a surfactant templating method, with cetyltrimethylammonium bromide (CTAB) micelles acting as template for HAp growth. The effects of the sintering temperature on the morphological and crystallographic characteristics and on chemical composition of the "as-prepared" structures are discussed. The experimental results show that low heat-treatment temperatures are preferred in order to obtain high quality nanorods, with diameters ranging between 20 nm and 50 nm. High heat-treatment temperatures enhance the thermal decomposition of HAp into other calcium phosphate compounds, and the sintering of particles into micrometer ball-like structures. The stability of aqueous suspensions of HAp nanorods is also discussed.**

I. INTRODUCTION

In the last years, an intensive research on nanostructured materials, processed in different shapes like tubes, spheres, rods, and ellipsoids, has been carried out [1]. Among them, nanoparticles have drawn a much larger interest. Due to their dimensions, high surface/volume ratio and to the possibility of being functionalized, they stand as very adequate candidates for many applications in Medicine. In this regard, they have already been tested as drug carriers, cell growth templates, and magnetic resonance contrast agents [2-4].

In order to prevent rejection reactions, nanostructured materials used for medical applications must be biocompatible, non-citotoxic and biodegradable. Since those, which are based on calcium phosphates are chemically similar to calcified biological tissues, they exhibit remarkable biocompatibility, and are widely used in orthopedics and dentistry as both prostheses coatings and bone fillers [5]. Currently, several calcium phosphates are being often used, such as hydroxyapatite (HAp), tricalcium phosphate (α and β-TCP) and dicalcium phosphate (BCP). However, since HAp is the only one stable at physiological conditions, it is thus the most commonly used for medical applications [5].

Several methods are available for synthesizing HAp-based nanostructured materials, like sol-gel, reverse microemulsion, hydrothermal, microwave-hydrothermal, solid-state reaction, and precipitation [6]. The latter one is the most often used process, as it is a simple, low cost, and suitable method for industrial production [7]. An often-used method to processing nanostructured materials involves the use of surfactants as templates [8,9]. These molecules are usually made of two parts, one hydrophilic and another hydrophobic. In a polar solvent, these molecules join together in such a way that the hydrophilic part faces outwards, thus being in contact with the solvent, while the hydrophobic part points inwards, forming a molecular structure, called micelle [10].

Currently, the most often technique used to synthesize HAp-based nanostructured materials is the cationic surfactant template method [11,12]. By controlling the processing parameters, like temperature and reactants concentration, it is possible to modify micelle growth, yielding nanostructures with very different properties [13].

A very special class of these materials is the mesoporous materials group. The synthesis of this particular type of materials was firstly reported in 1992 [14, 15]. They are commonly used as catalysts, absorbents, separators, host materials, templates, nanoshells for photothermal cancer therapy, and tissue-welding materials [8, 11, 16].

Despite the very intensive research that has been done to sort out the most suitable chemical routes for processing nonporous HAp nanostructures, no systematic study has yet been done in order to figure out the effect of processing parameters on their morphology, crystalline structure, and chemical composition.

This work is aimed at synthesizing HAp nanorods, through the cationic surfactant template method, by using cetyltrimethilammonium bromide (CTAB) micelles, acting as templates for HAp growth. CTAB was chosen since it yields stable micelles, whose shape and size can be easily changed by altering both concentration and temperature of the solution. One of the main purposes of this work is to correlate morphology, crystalline structure, and chemical composition of the "as-prepared" nanorods with sintering temperature. Moreover, the stability of aqueous suspension of HAp nanorods is also studied. This is a major issue as their further functionalization definitely depends on stability.

## II. EXPERIMENTAL DETAILS

A. Synthesis method

The experimental method for processing nanorods of HAp by using the cationic surfactant method is based on a method previously described [11]. In this work we have used 4,182 g of $K_2HPO_4$ (*Merck*, 99%) and 4,900 g of $CaCl_2.H_2O$ (*Merck*, 99%). The $PO_4^{3-}$:CTAB = 1: 1 ratio was kept unchanged throughout the work. 100 mL aqueous solution of $K_2HPO_4.3H_2O$ and CTAB (*Merck*, 99%) was prepared, with pH adjusted to 12, by adding 0.01 M NaOH solution. As CTAB does not dissolve easily, the solution had to be sonicated. Consecutively, 60 mL of a 0,074g.cm$^{-3}$ solution of $CaCl_2$ were added dropwise to the CTAB-phosphate solution. This procedure was chosen, as CTAB solutions polymerize very quickly, turning into a gel. This kind of solution could easily obstruct a burette stopcock. So, $CaCl_2$ solution is added instead, while the CTAB-phosphate solution is being stirred. At this point, a milky suspension is obtained, which was then heated at 83 +/- 2 °C, for 24h, in a refluxing system. Precipitates were washed 6 times with de-ionized water, in order to remove excess ions, contaminants and unreacted CTAB. Finally, the powder was dried for 24h, and sintered at different fixed temperatures between 500 ºC and 900 ºC for 6h.

B. Characterization techniques

Several experimental techniques were used in order to study the morphology, chemical composition, crystallographic properties, and stability of the aqueous suspensions of the "as-prepared" nanorods.

The unpolarized micro-sampling Raman spectra of the HAp nanorods were recorded in the backscattering geometry, at room temperature, by using an Olympus BH2 UMA microscope, with a 50x magnification lens. The 514.53 nm polarized line of a Ar+ laser was used for excitation, with an incident power of about 150 mW impinging on the sample. The scattered light was analyzed using a T64000 Jobin-Yvon spectrometer operating in the double subtractive mode, and equipped with a $LN_2$ cooled CCD. Identical conditions were kept unchanged throughout all of the measurements. The spectral slit width and the spatial resolution were about 1.5 $cm^{-1}$ and 1μm, respectively.

Both crystal structure and phase purity were checked through x-ray powder diffraction. The x-ray diffraction patterns were obtained in an ENRAF NONIUS diffractometer, in a Debye-Scherrer configuration, using monochromatic Cu kα radiation (λ = 1.540598Å). A INEL-CPS120 detector was used in the diffractometer, with angular resolution of ~0.02° for 2θ. The samples were crushed in a mortar, and then sieved in a strainer (pore size ≈ 68 μm) to obtain uniform powder dimensions. The Rietveld method was used for spectra analysis. This procedure enabled us to obtain the percentage of spurious phosphate phases, besides HAp.

The morphology of the nanorods was studied by using scanning electron microscopy (SEM). SEM images were obtained using a field emission JEOL JSM 6301F microscope, with a resolution of 1.2 nm, in secondary electrons mode. Transmission electron microscopy (TEM) analysis was performed with a Leica LEO 906E microscope, operating at 120 kV, and a resolution of ~ 0,344 nm.

The stability of the aqueous HAp suspensions was studied by measuring the zeta potencial, as a function of the chemical composition. In this work, we used 0.1 M solutions of NaCl, KCl and $KNO_3$, at physiological pH value (~7.4). A Zeta Sizer Nano ZS from Malvern was used for zeta potential measurements. It operates with a He-Ne laser (623 nm), and is suitable for zeta potential measurements of particles in the range 5 nm – 10 μm.

## III. RESULTS AND DISCUSSION

A. Morphological characterization.

Figures 1(a), (b) and (c) show SEM images of samples heat-treated at 550 ºC, 700 ºC, and 900 ºC, respectively. Fig. 1 (d) shows a lower magnification view of the SEM image for the sample treated at 900 ºC. The powder obtained after heat-treatment at temperatures between 550 ºC and 700 ºC is composed by well defined, randomly oriented elongated nanorods. As the temperature increases towards 700 ºC, elongated shapes of the nanorods are maintained, but for temperatures above 700 ºC, sphere-like structures are observed.

Figure 2 exhibits the mean diameter value of nanorods versus heat-treatment temperature. For temperatures between 550 ºC and 650 ºC (temperature range (a)), the average diameter of the nanorods takes values in a rather narrow interval that increases with increasing temperature. Though the average diameter shows a slightly decrease at 700 ºC, this is actually not a significant issue. On the other hand, for sintering temperatures above 700 ºC (range (b)), the mean diameter strongly increases as the temperature rises, wherein the spread of the diameter values becomes larger. The large spread may probably be due to the partial or incipient fusion of adjacent nanorods surfaces, as it may be observed in Figure 1(d).

Nanorods length ranges from 500 nm to ~2 μm, but no correlation between length and sintering temperature could be established.

In order to look for pores, we have performed TEM analysis of the HAp nanorods. Figures 3(a) and (b) show TEM images, obtained for the samples heat-treated at 550 ºC and 700 ºC, respectively. TEM images show the formation of pores-free compact structures. Their absence can be likely due to the collapse of the structure during the sintering process.

B. Structural and chemical phase characterization.

Figure 4 shows the Raman spectrum of nanophased HAp microaggregates, from a commercial sample, provided by Fluidinova S.A (Portugal), recorded in the spectral range 200 cm$^{-1}$ - 1200 cm$^{-1}$, at room temperature. As this sample presents high chemical purity, the recorded Raman spectrum will be used for comparison purposes with the ones obtained from our samples.

The Raman spectrum presented in Figure 5 is dominated by bands arising from the internal vibrations of the phosphate ion (PO$_4^{3-}$), as it was previously reported [17]. The free PO$_4^{3-}$ ion with T$_d$ symmetry has nine normal vibration modes: the totally symmetric stretching mode $\nu_1$, the double degenerated $\nu_2$, the triple degenerated asymmetric stretching mode $\nu_3$, and the triple degenerate asymmetric bending mode $\nu_4$. In a complex crystal, due to the lower symmetry of the local crystal field, both shift and splitting of the Raman bands are expected, due to molecular distortions. So, the profile of the Raman spectra can give relevant information concerning the crystal structure. In HAp crystals, with hexagonal symmetry, the degeneracy of the modes is entirely lifted. Taking into account the relevant spectroscopic information concerning the frequencies of the PO$_4^{3-}$ ions internal vibration, the following mode assignment of the Raman bands observed in Figure 4 is presented.

The stronger band located at 950 cm$^{-1}$ is assigned to the symmetric stretching mode $\nu_1$ (yielding the highest Raman intensity), the duplet observed between 400 – 450 cm$^{-1}$ to the symmetric bending mode $\nu_2$, the well resolved triplet between 1000 – 1150 cm$^{-1}$ to the asymmetric stretching mode $\nu_3$, and the triplet between 550 – 650 cm$^{-1}$ to the asymmetric bending mode $\nu_4$ [17].

The Raman band arising from the O-H stretching mode, usually located at 3576 cm$^{-1}$, is not well observed. This fact has been reported in other spectroscopic studies on HAp nanorods, and it has been associated with the low degree of crystallinity of the samples [18].

Figure 5 exhibits the Raman spectra of the HAp particles, obtained after sintering of the "as-precipitated" powder, both at different fixed temperatures and different locations in the sample.

The Raman spectra of the nanorods heat-treated at 550 ºC (see Figure 5(a)), exhibit the typical signature of the HAp pure phase, independently of the location, they were recorded from. These results clearly evidence both homogeneity and phase purity of the sample heat-treated at 550 ºC. As the sintering temperature increased, Raman spectra exhibit new Raman bands, some of them dependent on the recording location of the sample, providing that the samples are not chemically homogeneous. This result should be understood in the scope of thermal decomposition of HAp, into other calcium phosphates compounds, enhanced by the sintering temperature. This assumption is based on the progressive vanishing of the Raman bands associated with HAp molecules, along with the raising of new bands, as the temperature increases. In the spectrum displayed in Figure 5(b), the characteristic $v_1$ band assigned to the symmetric stretching mode of the $PO_4^{3-}$ ion is still visible. A sharp band located at 1050 cm$^{-1}$ is also observed, which is typical of carbonate apatite, arising from the coupling between the $v_1$ mode of $CO_3^{2-}$ with the $v_3$ mode of $PO_4^{3-}$ [19]. Although, we removed the dissolved $CO_2$ from the water through boiling, before using it, it is likely that some $CO_2$ has been re-dissolved during the solution preparation. Thus, since the samples were prepared and sintered under atmospheric conditions, some carbonation of the samples is expected during the HAp precipitation and sinterization process.

The thermal decomposition process is stronger for higher sintering temperatures. The Raman bands observed in Figure 5(c), correlates very well with β-TCP phase, recognized by the presence of the typical bands at 950 cm$^{-1}$ and 970 cm$^{-1}$ [20]. These bands arise from the significant differences in the inter-tetrahedral P-O band length for different nonequivalent $PO_4^{3-}$ ion of the β-TCP structure [20]. Other typical β-TCP bands are observed in the 400 – 500 cm$^{-1}$ and 550 – 650 cm$^{-1}$ frequency range, confirming the formation of β-TCP. The characteristic peak at 1050 cm$^{-1}$ evidences the presence of carbonated apatite.

Thus, high sintering temperatures induce the formation of other phosphate phases, namely β-TCP and carbonated apatite, apparently due to the thermal decomposition of HAp. It may then be concluded that samples become chemically heterogeneous as sintering temperature is increased above 550 ºC.

In order to figure out the crystal structure of the "as-prepared" powder, and to confirm the existence of spurious calcium phosphate phases, powder diffraction spectra were recorded at room temperature. Figures 6(a) and 6(b) show the x-ray diffraction spectra for the samples heat-treated at 550 ºC and 800 ºC, respectively. Both x-ray spectra exhibit a high and broad background at small angles. Due to the presence of other chemical phases in the samples, differences in peaks position and intensity can be observed. Detailed Rietveld analysis reveals the existence of spurious phosphate phases, namely β-TCP and carbonated apatite, in the sample sintered at the highest temperature, in good agreement with the Raman spectroscopy results. The sample heat-treated at 550 ºC exhibits predominantly the hexagonal HAp phase.

C. Suspension stability

Whenever particles are dispersed in a solution, a suspension is formed. Generally, stable suspensions are preferred for particles functionalization, since individual particles exhibit a higher effective surface. Moreover, their stability prevents particles to attract each other, and thus the formation of heavy aggregates, which could lead to surplus, disadvantageous precipitates.

A way to evaluate whether a particular suspension is stable or not, is to ascertain its zeta potential. A suspension is considered unstable if its zeta potential presents values in the range -30 mV to 30 mV [17]. Therefore, we carried out a study of the stability of "as-prepared" HAp nanorods, treated at 550 ºC, in an aqueous solution of 0.01M of NaCl, KCl and $KNO_3$, at physiological pH (~7.4). Table I presents the values of the zeta potential for the suspensions referred to above.

As it can be observed, HAp-NaCl and HAp-$KNO_3$ suspensions are not stable, and thus particles will tend to aggregate creating a precipitate. Even though, some applications could be figured out, as the precipitated clusters can be further used for functionalization purposes, as in the case of nanosponges.

From the results of Table I, we can assert that the HAp - KCl suspension is stable and thus can be further used for functionalization purposes. Though the other two suspensions are unstable, they can be used for HAp microsponges processing.

IV. CONCLUSIONS

This work addresses the characterization of elongated nanorods of HAp, obtained from a cationic surfactant template method, sintered at different temperatures. A clear dependence of chemical homogeneity, structure and morphology on heat-treatment temperature is observed.

Sintering temperature has a major effect on the morphology and chemical composition, inducing thermal decomposition of HAp into β-TCP and carbonated apatite. Moreover, it also yields the combination of the nanorods into micrometric spheric-like structures.

The samples obtained after heat-treatment at temperatures in the range 550 ºC-700 ºC, exhibit very well defined elongated and compact nanorods, with diameters ranging from 20 nm to 50 nm, slightly increasing with increasing temperature. The samples obtained after sintering at 550 ºC exhibit mainly the hexagonal HAp phase, with high chemical homogeneity, and small dispersion of the particles diameter.

As the sintering temperature increases, other phosphate phases are formed, namely β-TCP, though in small amounts. Moreover, aggregation of nanorods occurs, whereas their diameter increases, eventually reaching a much larger value. For the samples treated at 900 ºC, no individual particles could be observed. Instead, sphere-like microstructures are formed, exhibiting strong chemical heterogeneities.

The stability of aqueous suspensions prepared from HAp nanorods that were obtained after treatment at 550 ºC, was studied through the measurement of their zeta potential. Both HAp-NaCl and HAp-KNO$_3$ suspensions were confirmed as unstable. HAp-KCl suspension proved to be stable, which provides to be adequate for further functionalization processing. Though the other two suspensions are not useful for this purpose, they could be excellent candidates for processing microspongeous HAp materials.

**FIGURE CAPTIONS**

**Figure 1.** SEM images of samples sintered at: a) 550 ºC, b) 700 ºC and c-d) 900 ºC. Figure d) shows a lower magnification view of the sample sintered at 900 ºC.

**Figure 2.** Mean diameter values (dots) versus sintering temperature. The spreads in the diameter values are shown as yy error bars.

**Figure 3.** TEM images of HAp calcinated at (a) 550 ºC and (b) 700 ºC.

**Figure 4.** Raman spectra of commercial Hap powder (NanoXIM- Fluidinova.S.A).

**Figure 5.** Raman spectra for HAp heat-treated at different temperatures: (a) 550 ºC, (b) 700 ºC and (c) 900 ºC.

**Figure 6.** XRD pattern of HAp samples: a) 550 ºC and b) 800 ºC.

**TABLE CAPTIONS**

**Table I.** Zeta potentials for HAp sintered at 550 ºC.

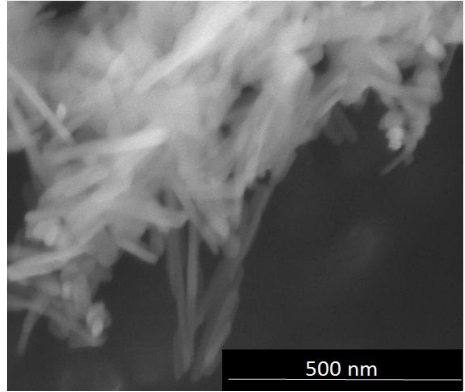

Figure 1(a)

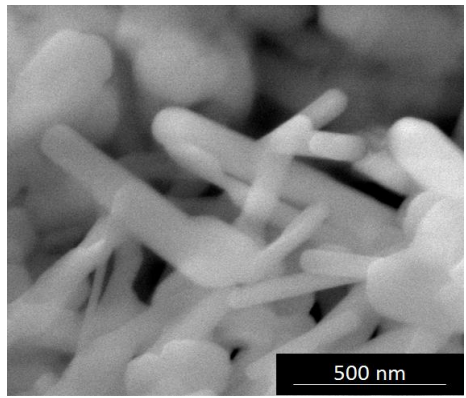

Figure 1(b)

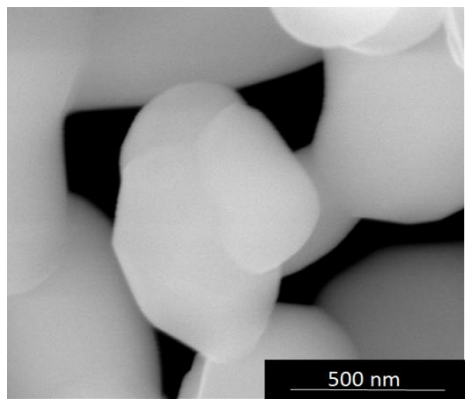

Figure 1(c)

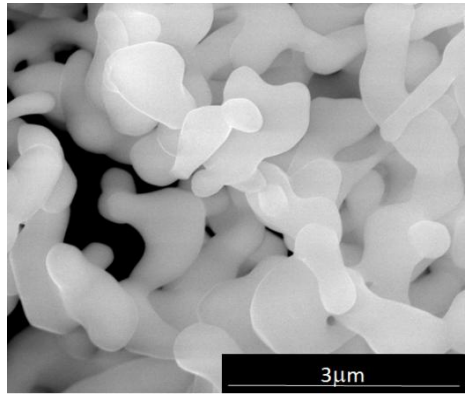

Figure 1(d)

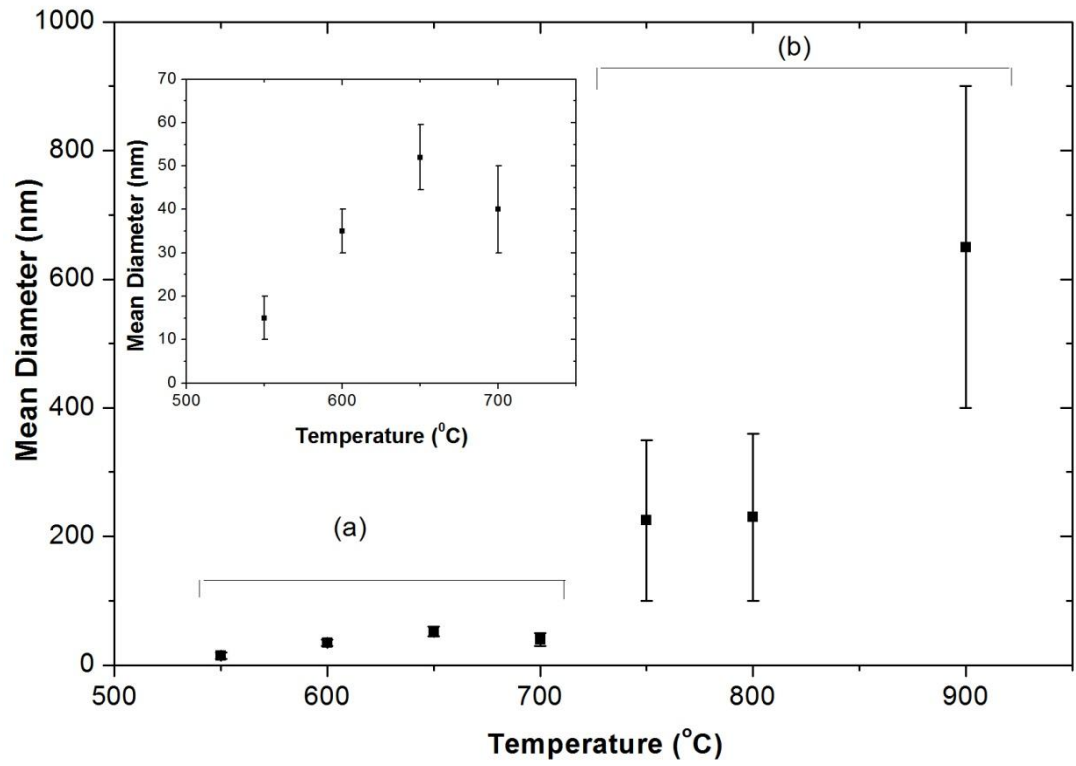

Figure 2

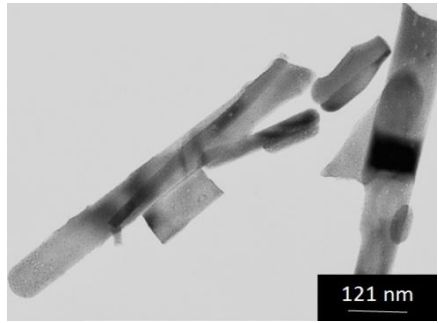

Figure 3(a)

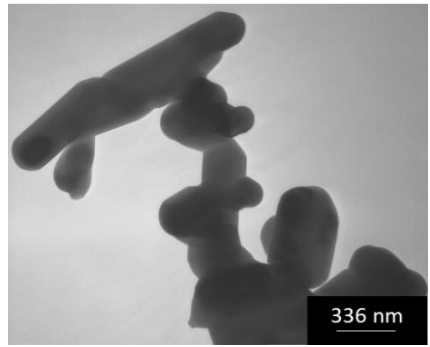

Figure 3(b)

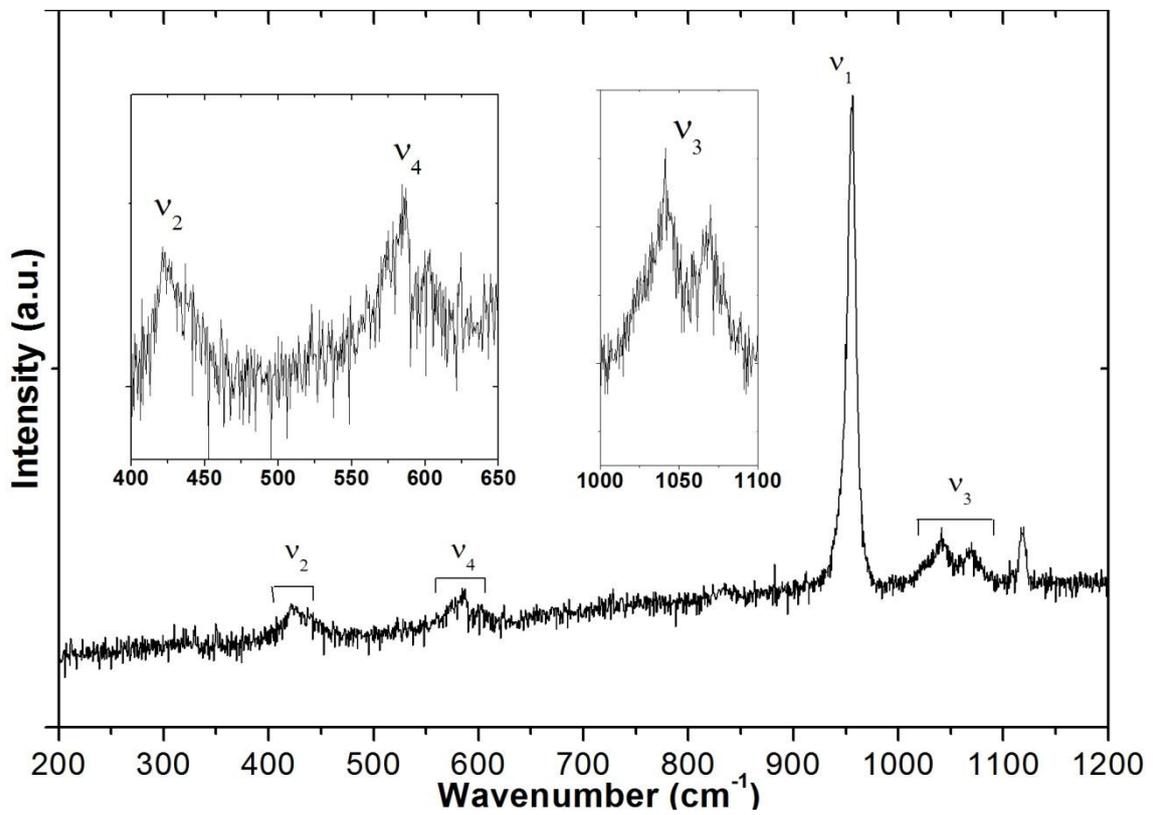

Figure 4

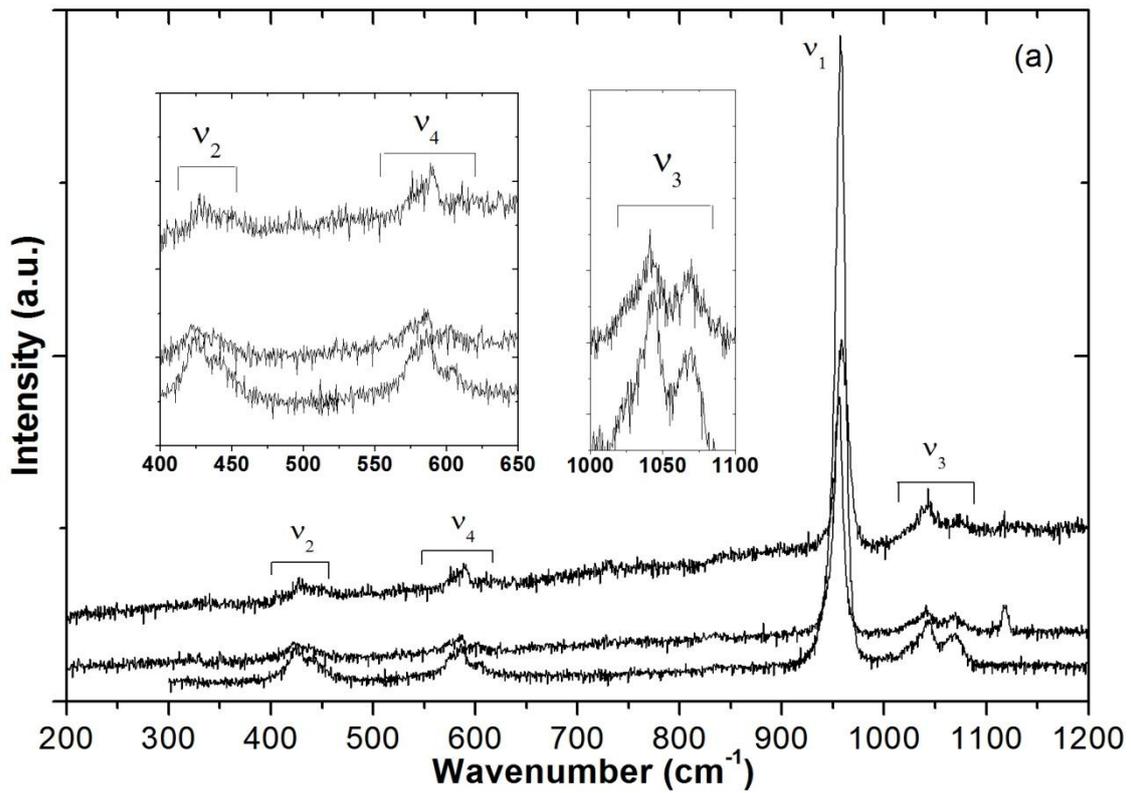

Figure 5(a)

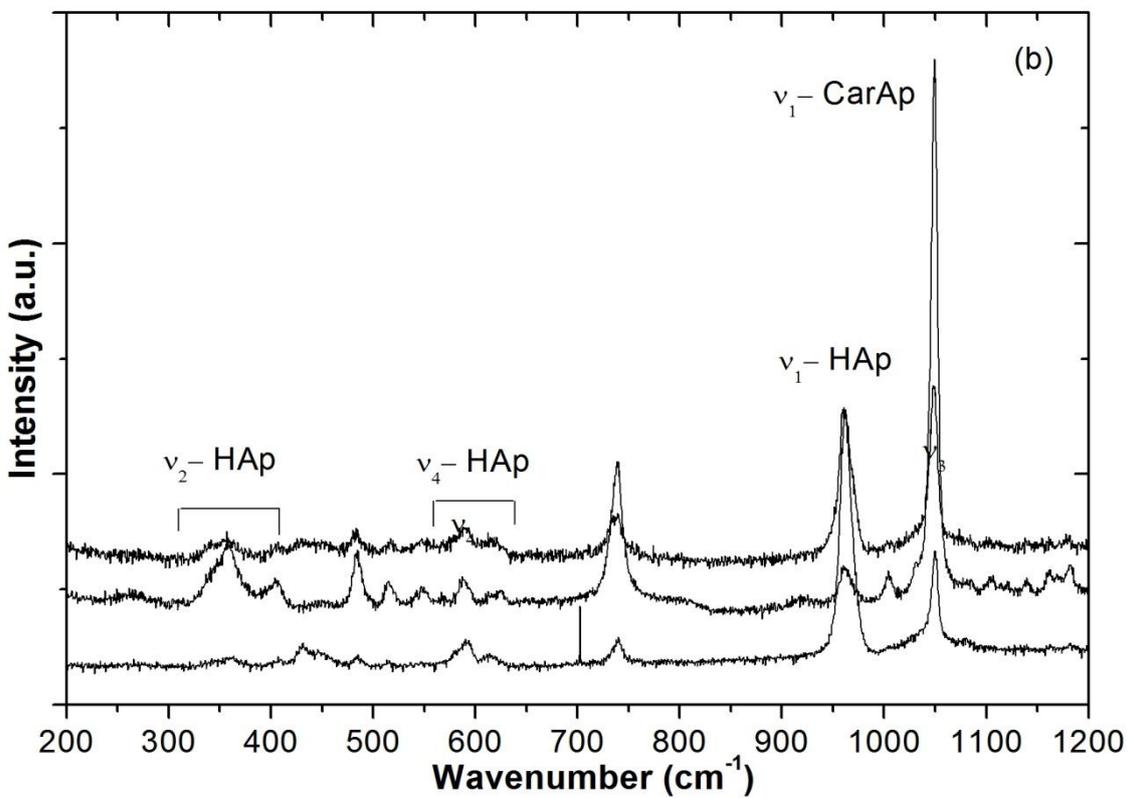

Figure 5(b)

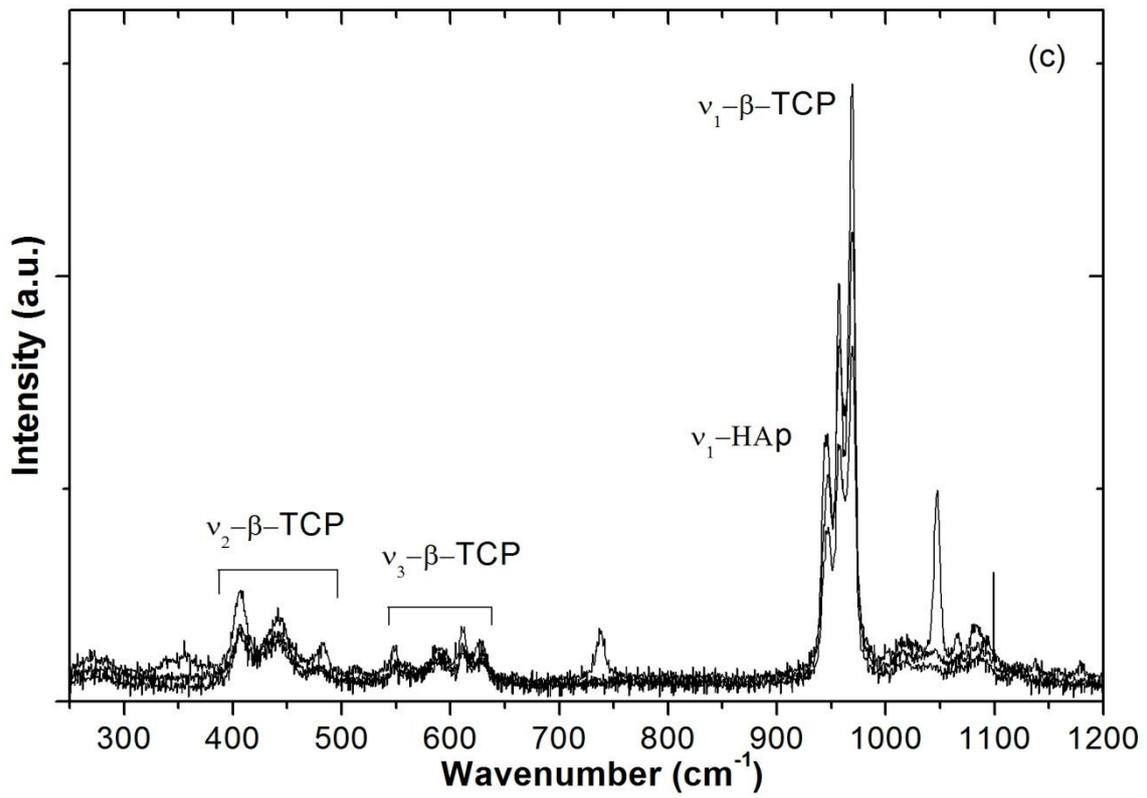

Figure 5(c)

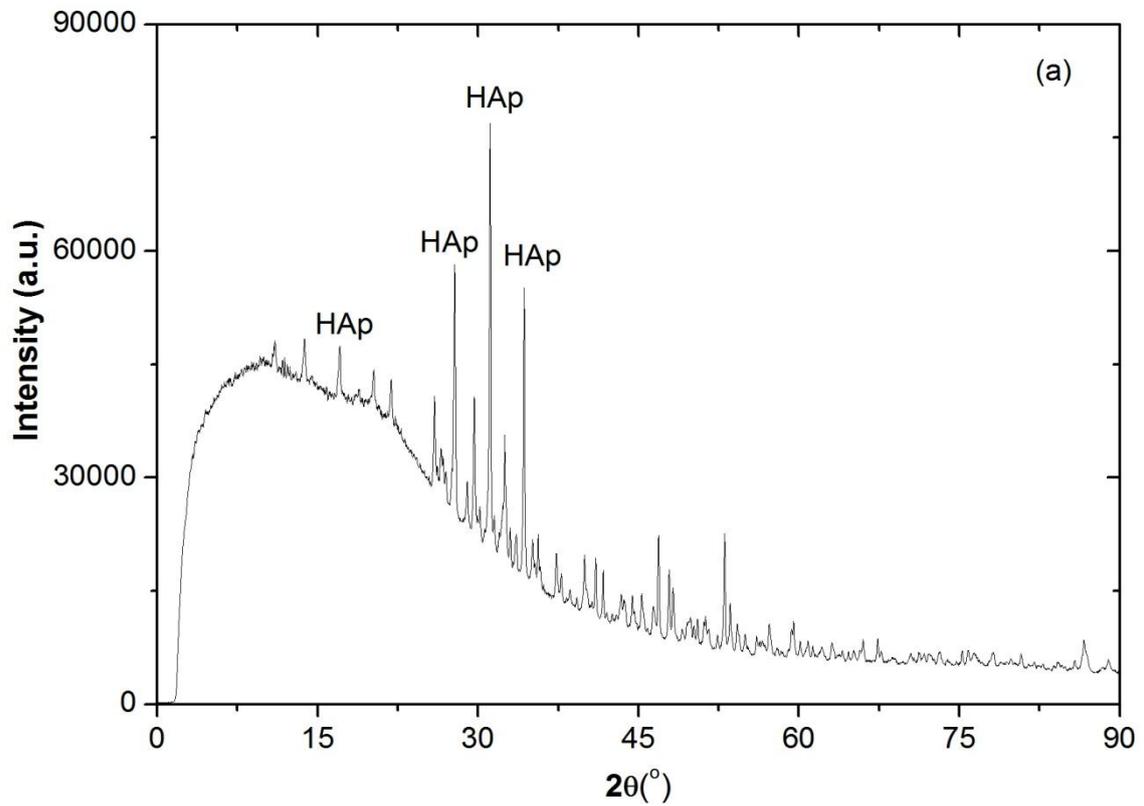

Figure 6(a)

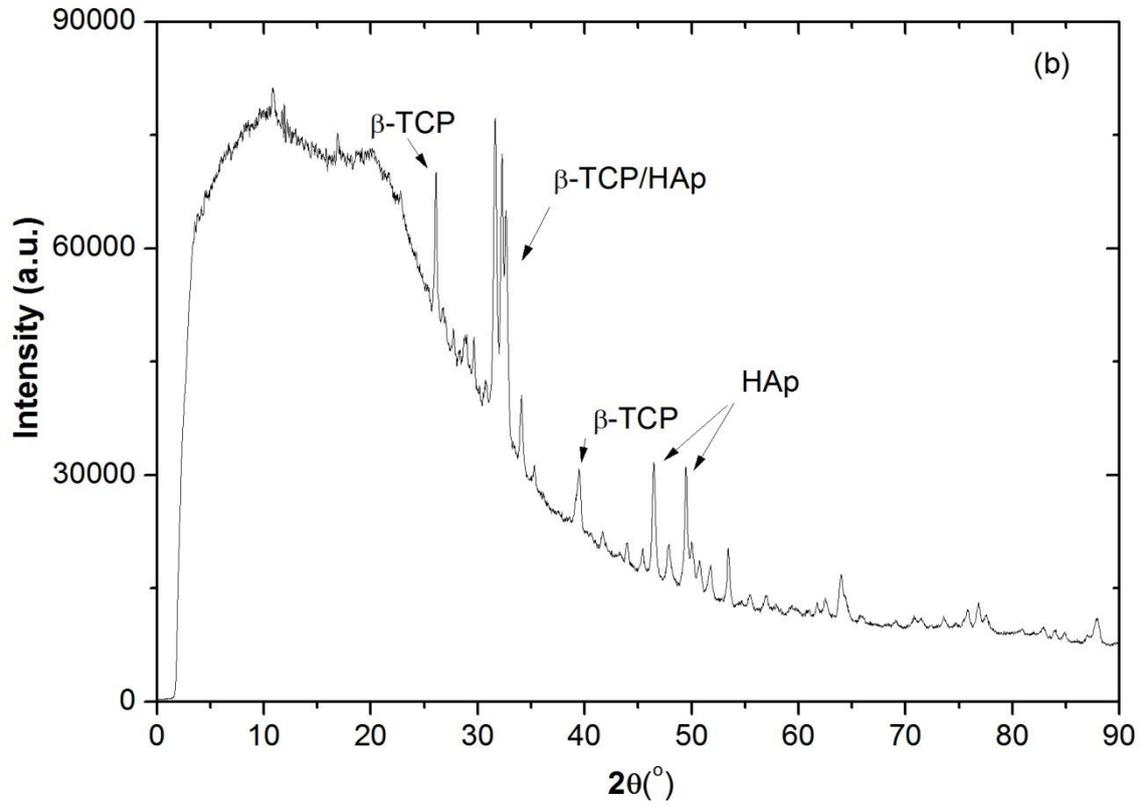

Figure 6(b)

| Solution | Zeta potential (mV) |
|---|---|
| 0.01 M NaCl | -9.18±0.40 |
| 0.01 M KCl | -32.18±1.15 |
| 0.01 M $KNO_3$ | -8.17±1.71 |

Table I